\documentclass[superscriptaddress,twocolumn,showpacs,showkeys,amsmath,prb,final]{revtex4}
\usepackage{graphicx}% Include figure files
\usepackage{dcolumn}% Align table columns on decimal point
\usepackage[sort&compress]{natbib}
\usepackage{subfigure}
\usepackage{ifpdf}
\usepackage{bm}
\ifpdf
\usepackage[pdftex,
        colorlinks=true,
        pdftitle={ Effect of transport of growing nanoparticles on
 ccrf discharge dynamics.},
        pdfauthor={ I.V. Schweigert },
        pdfsubject={},
        pdfkeywords={nanoparticle transport, gas discharge},
        baseurl={},
        pdfpagemode=UseNone,
        bookmarksopen=true
        pdfoutput=1
        ]{hyperref}
\fi

\begin{document}

\title{Theoretical and experimental study of ion flux formation in an
asymmetric high-frequency capacitive discharge}
\author{I.V. Schweigert}
\affiliation{Khristianovich Institute of Theoretical and Applied Mechanics,
Siberian Branch, Russian Academy of Sciences, Novosibirsk 630090, Russia}
\email{ischweig@itam.nsc.ru}
\author{D.A. Ariskin}
\affiliation{Khristianovich Institute of Theoretical and Applied Mechanics,
Siberian Branch, Russian Academy of Sciences, Novosibirsk 630090, Russia}
\author{T.V. Chernoiziumskaya}
\affiliation{St.-Petersburg State Polytechnical University, Russia}
\author{A.S. Smirnov}
\affiliation{St.-Petersburg State Polytechnical University, Russia}

\date{\today}

\begin{abstract}
Parameters of a high frequency capacitive discharge in argon in
axially symmetric chambers of different geometries are studied in experiments and
by means of two-dimensional kinetic modeling by the Particle-in-Cell method.
It is demonstrated that a change in the ratio of the areas of 
the driven and grounded
electrodes can substantially increase the ion energy on the electrode
practically without disturbing the plasma parameters. Particular attention
is paid to studying the self-bias voltage and the ion distribution function
on the electrode for gas pressures ranging from 15 to 70 mTorr. The results
of self-consistent calculations are in good agreement with experimental
data.
\end{abstract}
 \maketitle
%\keywords{asymmetrical gas discharge, ion flux}
%\pacs{52.25.-b, 52.25.Jm, 52.25.Dg}
%\maketitle

\section{Introduction}

Ion fluxes in sheaths adjacent to the electrodes generated by a 
high-frequency discharge are widely used in various plasma technologies. 
Ion bombardment results in anisotropy of etching processes and affects 
the growth rate and the structure of films grown in the discharge. The 
applied voltage drop in the sheaths is substantial, 
and a considerable
part of energy imparted to the discharge can be spent on accelerating ions. 

Processes affecting the sheath formation were studied
in detail by many researchers (see, e.g., \citep{1}). The majority of
theoretical and numerical works, however, are restricted to studying a
one-dimensional problem, whereas real discharge geometry is usually asymmetric.
In this case, the current density and, correspondingly, the drop of
voltage near one electrode is greater than the corresponding values in
the vicinity of the opposite electrode. Thus, there appears a constant
self-bias voltage, which appreciably increases the energy of ions
bombarding the electrode with the higher current density \citep{2}. 

The discharge is asymmetric because one of the electrode is connected to the
grounded walls of the discharge chamber, and some portion of the discharge
current is spent there. As such a discharge has a complicated two-dimensional
geometry, the distribution of currents can be calculated only with
two-dimensional numerical simulations. The use of the simple ratio of the
electrode areas (sheath capacities) in analyzing the experimental data  
\citep{2} is based on the assumption that the structure and size of the
sheaths near different surfaces are identical.  This assumption is not
 substantiated. Moreover, the sheath thickness and the distribution of the
charged particle concentrations in the sheath depends on the discharge 
 operation mode determined, among other parameters, by the current density
\citep{1,3}. The purpose of the present work is a numerical 
and experimental study of the ion flux formation in a low-pressure asymmetric 
high-frequency capacitive (HFC) discharge.

The paper is arranged as follows. The setup used in the experiments
is described in Sec. \ref{setup}. The kinetic model for
the two-dimensional description of the HFC discharge is given in Sec.
\ref{model}. The plasma parameters in chambers with different
geometries are compared in Sec. \ref{plasma}. The ion flux onto
the electrode is analyzed in Sec.  \ref{ionflux}. The calculated
results for the self-bias voltage and plasma potential, and also  the
distribution function of the ion fluxfor different gas pressures are
discussed in Sec. \ref{poten}. The conclusions are formulated
in Sec. \ref{conclusion}.

\section{Experimental setup}
\label{setup}

\begin{figure}[h]
\includegraphics[width=1.2\linewidth]{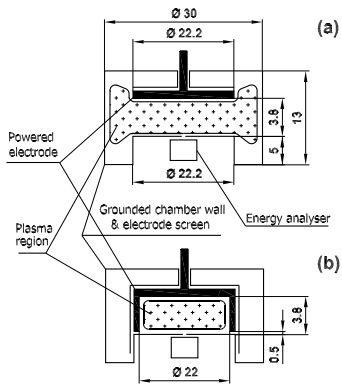}
\caption{Cylindrical gas-discharge chambers with different
ratios of the areas of the driven and grounded electrodes. 
Chamber A with top driven electrode and grounded bottom 
electrode and side wall (a) and
chamber B with top driven electrode and side wall and grounded bottom 
electrode (b).}
\label{Fig1}
\end{figure}

The electric parameters of a 13.56-MHz capacitive discharge in argon
were measured in the experiments performed in two chambers with different
configurations of the driven electrode. The gas-discharge chambers used
are schematically shown in Fig. \ref{Fig1}. 
Voltage is applied to the 
top electrode with an area $A_{rf}$, while the bottom electrode with an
area $A_0$ is grounded. In chamber A (see Fig. \ref{Fig1}(a)), the
effective area of the driven electrode is substantially smaller than the
grounded electrode area ($\delta S=A_{rf}/A_0 <1$), because the side
surface of the discharge chamber is also subjected to a zero potential.
With this ratio of the electrode areas $\delta S$, 
the discharge is visibly asymmetric at
low gas pressures, and the potential drop in the sheath adjacent to the
driven electrode prevails. The potential drop at the grounded electrode
sheath is small and depends only weakly on the gas pressure and on the 
input power. This configuration does not allow the flux
of high-energy ions from the discharge plasma to be studied, because
the ion energy analyzer is usually mounted in the high-vacuum chamber behind
the grounded electrode. Such a configuration of the driven electrode,
however, allows measurements of the radial distribution of the plasma
concentration and also the electron temperature by a moving Langmuir
probe. 

In chamber B, the driven electrode is supplemented with a cylindrical side 
part, which is shielded to prevent the breakdown on the side walls of the
chamber (see Fig. \ref{Fig1}(b)). In this configuration, it is only the
bottom electrode surface that is grounded, and the ratio of the areas of the
driven and grounded electrodes is $\delta S >1$. In chamber B, in 
asymmetric HFC discharge with a prevailing potential drop in the grounded
electrode sheath is formed, and it is possible to study the high-energy 
ion flux with the use of an energy analyzer. 

The gas pressure in the
experiments was varied from 6 to 70 mTorr.  The radius of chamber A is
$R$=15 cm, and its height is $H_0$=13 cm. The discharge glows 
between the electrodes with a radius of 11 cm; the distance between
the electrodes is $d$=3.8 cm. The radius of chamber B $R$=11 cm is equal 
to the electrode radius. 
 
To measure the ion flux distribution function over energy, there is
an orifice 0.1 cm in diameter at the center of the grounded bottom electrode;
this orifice connects the discharge chamber with the diagnostic chamber
located below. The diagnostic chamber is evacuated independent of discharge
chamber evacuation, and the pressure in the diagnostic chamber is lower than
1 mTorr. The diagnostic chamber contains a four-grid electrostatic energy
analyzer of the confining field. We used this analyzer to register
the energy distribution function of the ion flux moving from the discharge
plasma toward the grounded electrode. 

\section{Kinetic model}
\label{model}
The system of equations in a two-dimensional model of an HFC discharge
with cylindrical symmetry includes the kinetic equations for
electrons and ions (which are three-dimensional in terms of velocity and
two-dimensional in space) and Poisson's equation. The energy distribution 
function for electrons $f_e(\vec r,\vec v)$  and ions $f_i(\vec r,\vec v)$ 
are found from the Boltzmann equations 
\begin{equation}  \label{kine}
\frac {\partial f_e}{\partial t}+ \vec v_e\frac {\partial 
f_e}{\partial \vec r}
-\frac {e\vec E}{m}\frac {\partial f_e}{\partial \vec 
v_e}= 
J_e,\quad n_e=\int
f_ed\vec v_e,
\end{equation}
\begin{equation}  \label{kini}
\frac {\partial f_i}{\partial t}+ \vec v_i\frac {\partial 
f_i}{\partial \vec r}
+ \frac {e\vec E}{M}\frac {\partial f_i}{\partial \vec 
v_i} 
=J_i,\quad
n_i=\int f_id\vec v_i,
\end{equation}
where $v_e$, $v_i$, $n_e$, $n_i$, $m$, and $M$ are the electron and ion
velocities, concentrations, and masses, respectively; $J_e$ and $J_i$ 
are the collisional integrals for electrons and ions.

Knowing the energy distribution functions for electrons and ions,
we can calculate the mean energy of electrons and ions:
\begin{equation}  \label{edens}
\varepsilon_{e,i}(\vec r) = n_{e,i}^{-1}\int \frac{m_{e,i} 
v_{e,i}^2}{2} f_{e,i} d^3 v_{e,i}. \; 
\end{equation}

Poisson's equation describes the electric potential distribution
\begin{equation}  \label{Poisson}
\bigtriangleup \phi =4\pi e \left(n_e-\sum_{i=1}^N n_i \right),
\quad
\vec E=-\frac{\partial \phi}{\partial \vec r} \; .
\end{equation}
The boundary conditions for Poisson's equation are the voltage $U=0$ on the grounded
electrode and $U=U_0sin(\omega t)+U_{bias}$ on the driven electrode. The
self-bias voltage $U_{bias}$ is calculated from the condition of  a zero
total current onto the grounded surfaces and surfaces with applied voltage.

System (\ref{kine})-(\ref{Poisson}) is solved self-consistently by the
Particle-in-Cell  method with sampling of collisions by the Monte
Carlo method (PIC MCC) \citep{bird}. The HFC discharge operates in argon. The kinetics of
electrons includes elastic scattering of electrons on atoms, excitation
of metastable states, and ionization. Emission of secondary electrons
from the electrodes due to bombardment by ions with the secondary emission coefficient
$\gamma$ is also considered.

\section{Comparison of discharge parameters in chambers of different
geometries}
\label{plasma}
\begin{figure}[h]
\includegraphics[width=0.85\linewidth]{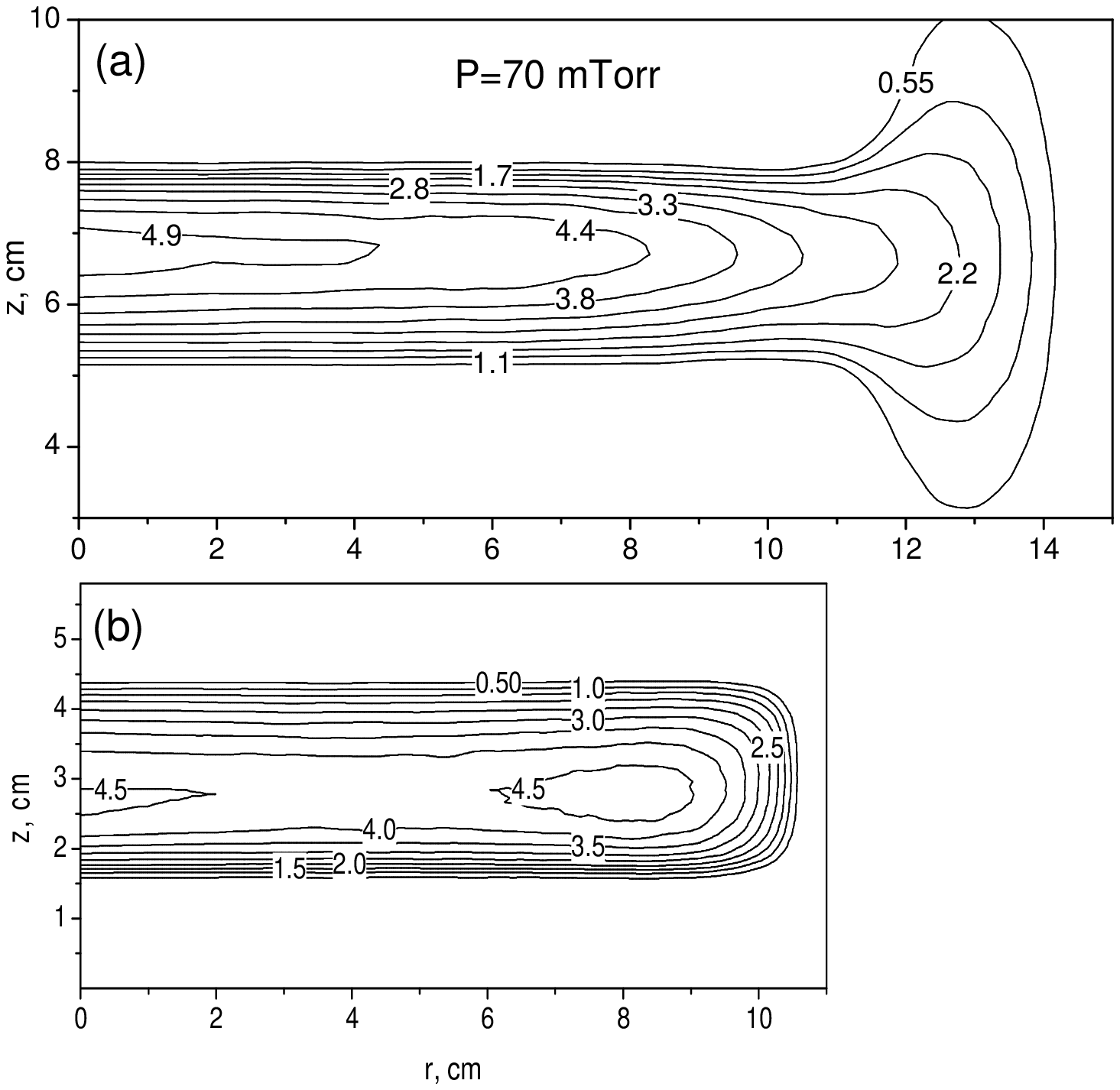}
\caption{Calculated distributions of the electron concentration in
chambers A (a) and B (b) for P=70 mTorr.
}
\label{den2D_70}
\end{figure}

Let us consider the plasma parameters obtained in the experiment and
in the self-consistent numerical solution of system (\ref{kine})-(\ref{Poisson}). 
The discharge operates in chambers of different geometries (see Fig. \ref{Fig1}) 
with a fixed power of 10 W, P=5, 30, and 70 mTorr, and $\gamma $=0.1. 
In chamber A, the electrode radius is $r_l$=11 cm, and the chamber radius is
$R$=15 cm. In chamber B, the bottom electrode radius is $r_l$=10.5 cm,
and the chamber radius is $R$=11 cm. These chambers have different ratios of 
the areas of the driven and grounded electrodes $\delta S$, because
the voltage is applied only to the top electrode in chamber A
($\delta S <1$) and to the top electrode and the side walls of the chamber
($\delta S >1$) in chamber B.
The calculated distributions of the electron concentration $n_e$ in the
chambers is shown in Fig. \ref {den2D_70} for $P=70$ mTorr.  
The electron 
concentration at the center of the discharge gap is $4.9\times 10^9$cm$^{-3}$ 
in chamber A and $4.5\times 10^9$cm$^{-3}$ in chamber B. In chamber A,
the concentration $n_e$ decreases with distance from the axis of symmetry
toward the side wall. In chamber B, there is a second peak of the plasma
concentration near the electrode edge owing to enhanced ionization.

\begin{figure}[h]
\includegraphics[width=0.85\linewidth]{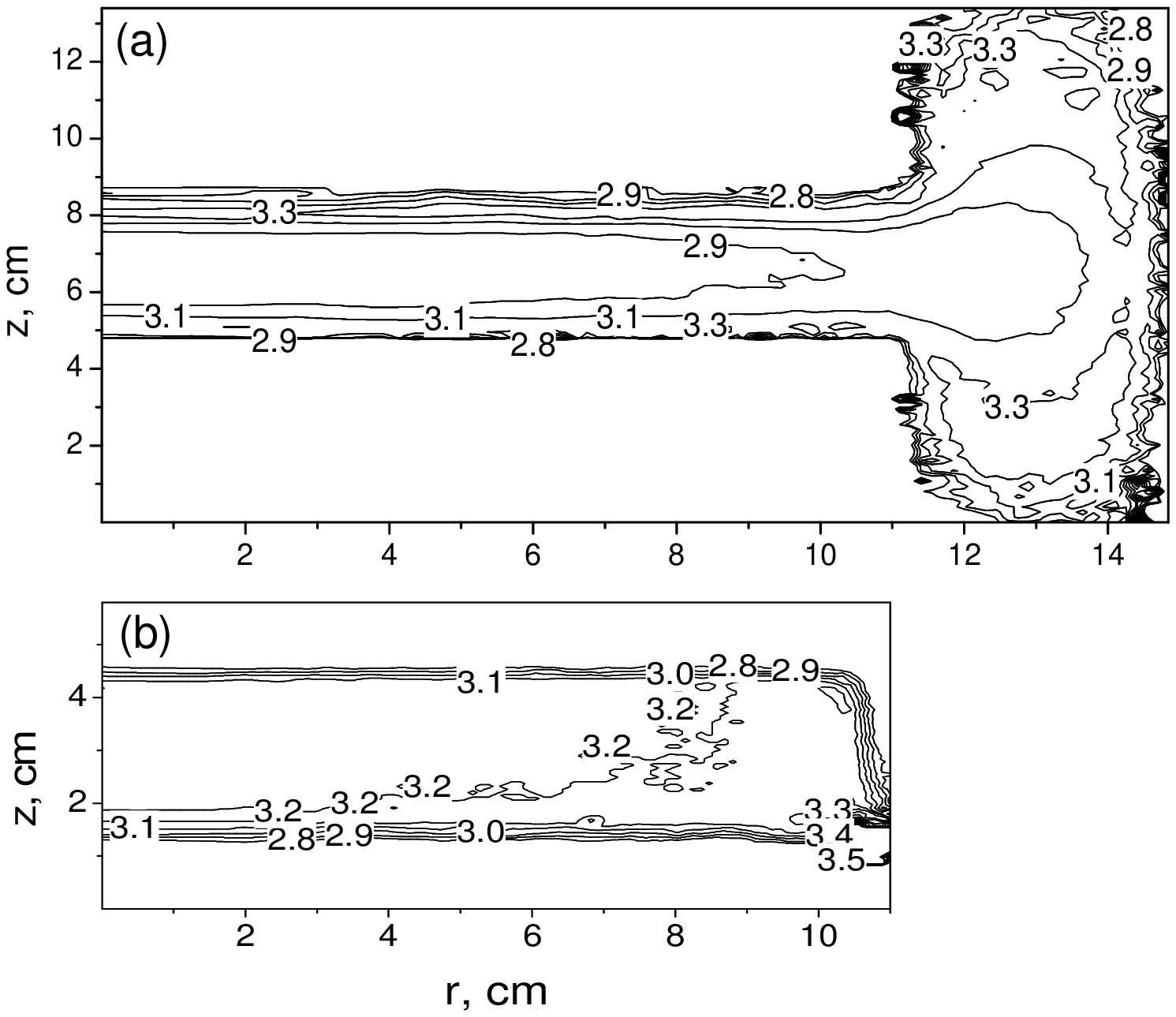}
\caption{Distributions of the mean electron energy for the same conditions
as in Fig. \ref {den2D_70}.}
\label{ene2D_70}
\end{figure}

Figure \ref {ene2D_70} shows the distribution of the electron energy
$\epsilon _e$. 
The energy of the electrons changes from 2.9 to 3.5 eV
over the volume of chambers A and B. Note that the maximum energy of
the electrons in chamber B is observed between the grounded bottom
electrode and the side walls, which is under voltage. The ionization rate here,
however, is not too high, because the gap length is smaller than the
characteristic length of ionization by electrons.

\begin{figure}[h]
\includegraphics[width=0.9\linewidth]{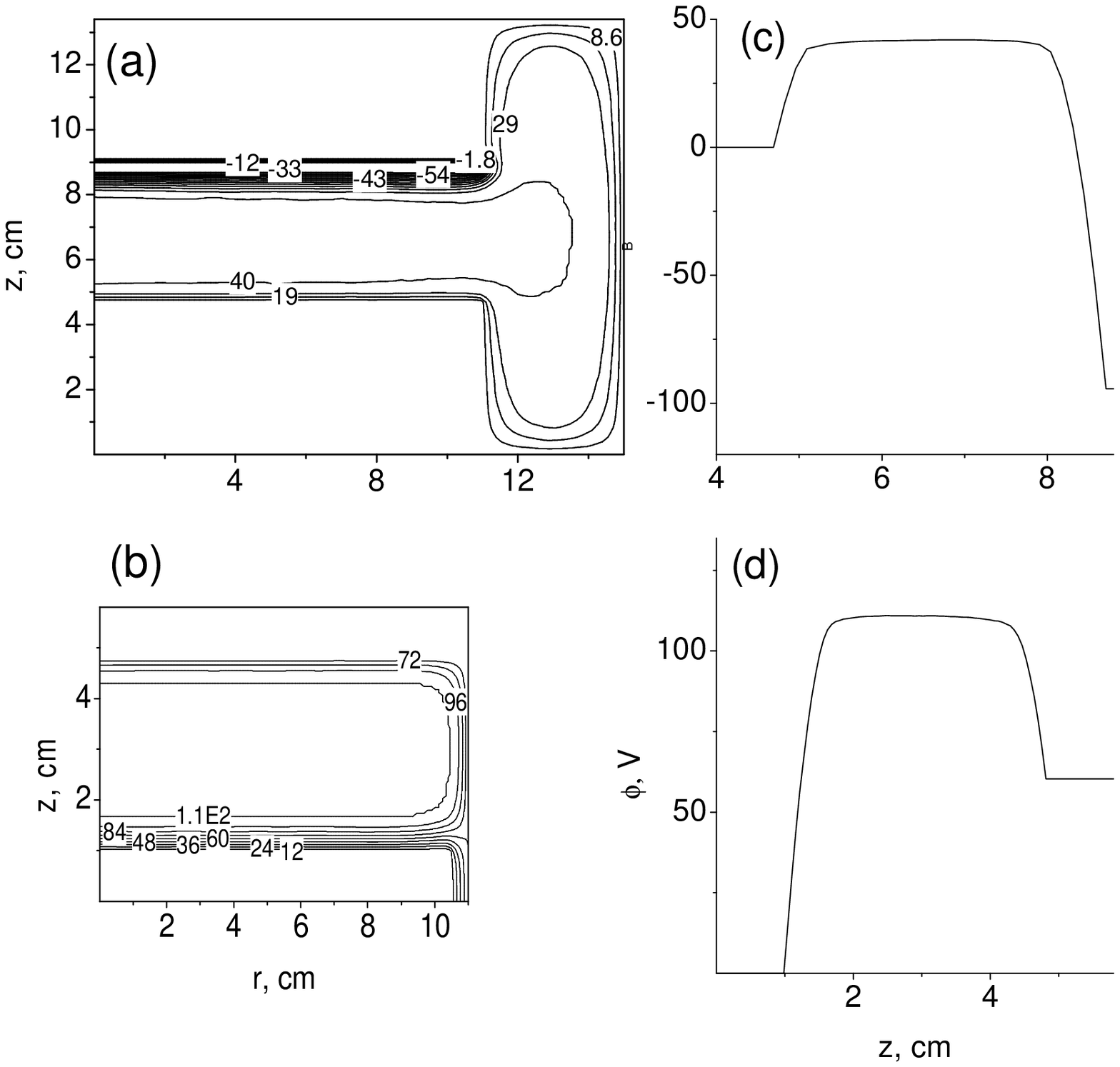}
\caption{Distributions of the electrical potential in chambers A (a,c) and 
B (b,d) for P=70 mTorr. The potential profiles on (c),(d) are shown for r=0.} 
\label{pot2D_70}
\end{figure}

The distribution of the electric potential $\phi$ in Fig. \ref {pot2D_70} 
demonstrates the appearance of the self-bias voltage induced by the difference
in the areas of the driven and grounded electrodes. 
Figures \ref {pot2D_70}(a),(c) 
show the distribution of  $\phi$ for chamber A, where the voltage is applied
to the top electrode with a smaller area; in this case, the self-bias
voltage is -100 V. Figures \ref {pot2D_70}(b),(d) refer to chamber B where
the grounded bottom electrode has a smaller area; the potential drop here is
110 V.

\begin{figure}[h]
\includegraphics[width=0.7\linewidth]{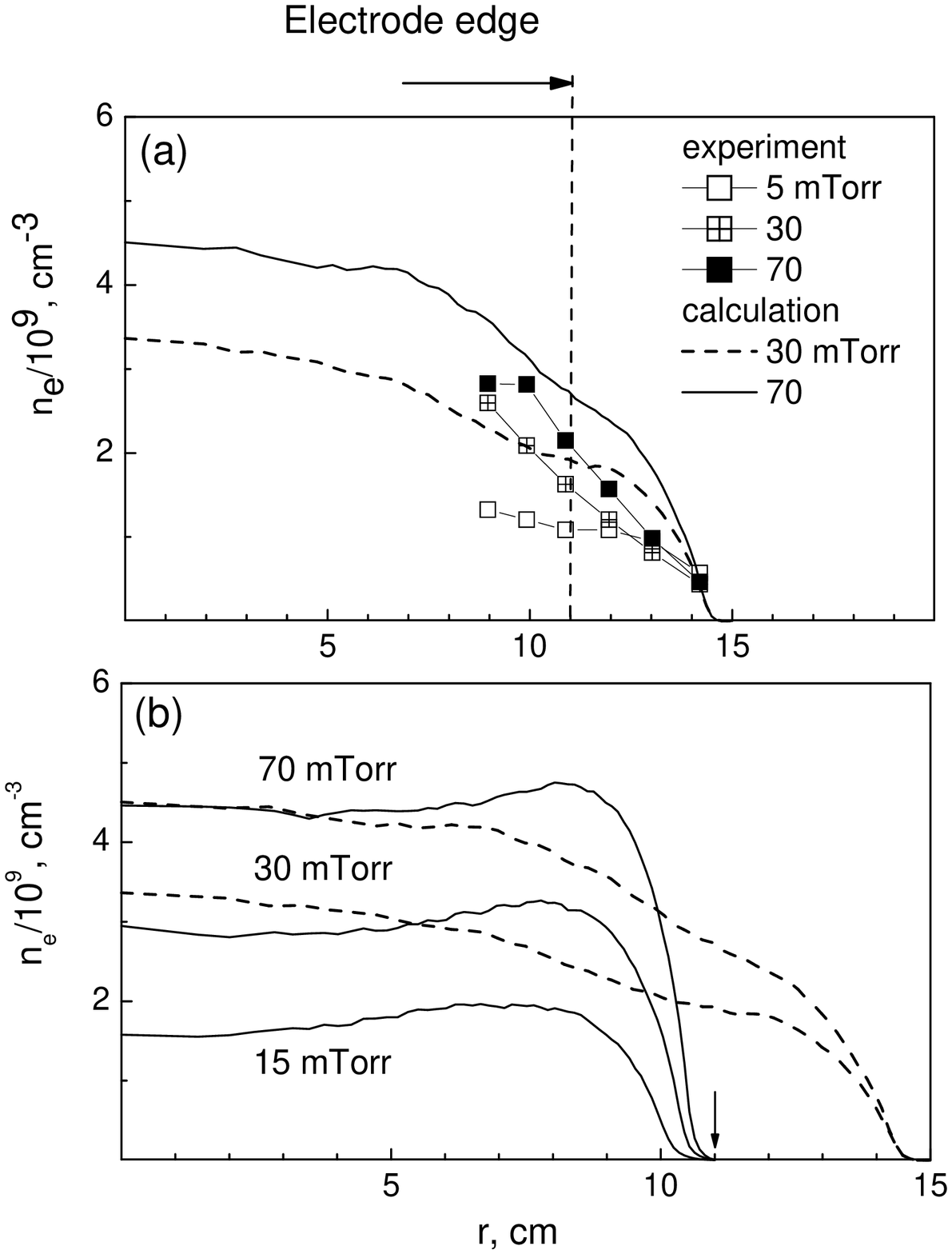}
\caption{Radial distributions of the electron concentration at the center
of the discharge gap (à) in chamber A, P=5, 30, and 70 mTorr (symbols, 
experiment), P=30 and 70 mTorr (curves, calculation)  and (b) in chambers
A and B,  P=15, 30, and 70 mTorr (calculation). The arrows indicate
the edge of the bottom electrode.}
\label{den_all}
\end{figure}

Let us compare the measured and calculated radial distributions of the electron
concentration and energy at the discharge gap center. Figure \ref {den_all}(à) 
shows the measured and calculated concentrations of electrons in chamber A
for different gas pressures. 
The calculated profiles of $n_e$ in chambers
A and B are plotted in Fig. \ref {den_all}(b). In chamber A, which has a greater
radius, the plasma concentration monotonically decreases toward the side wall.
An increase in the electron concentration near the edge of the bottom electrode
is observed in chamber B. As a whole, the change in the chamber geometry has
a weak effect on the plasma concentration for gas pressures ranging from
15 to 70 mTorr. With increasing pressure, the plasma concentration increases 
in both chambers from  1.8$\times 10^9$cm$^{-3}$ to 4.5$\times 10^9$cm$^{-3}$.

\begin{figure}[h]
\includegraphics[width=0.7\linewidth]{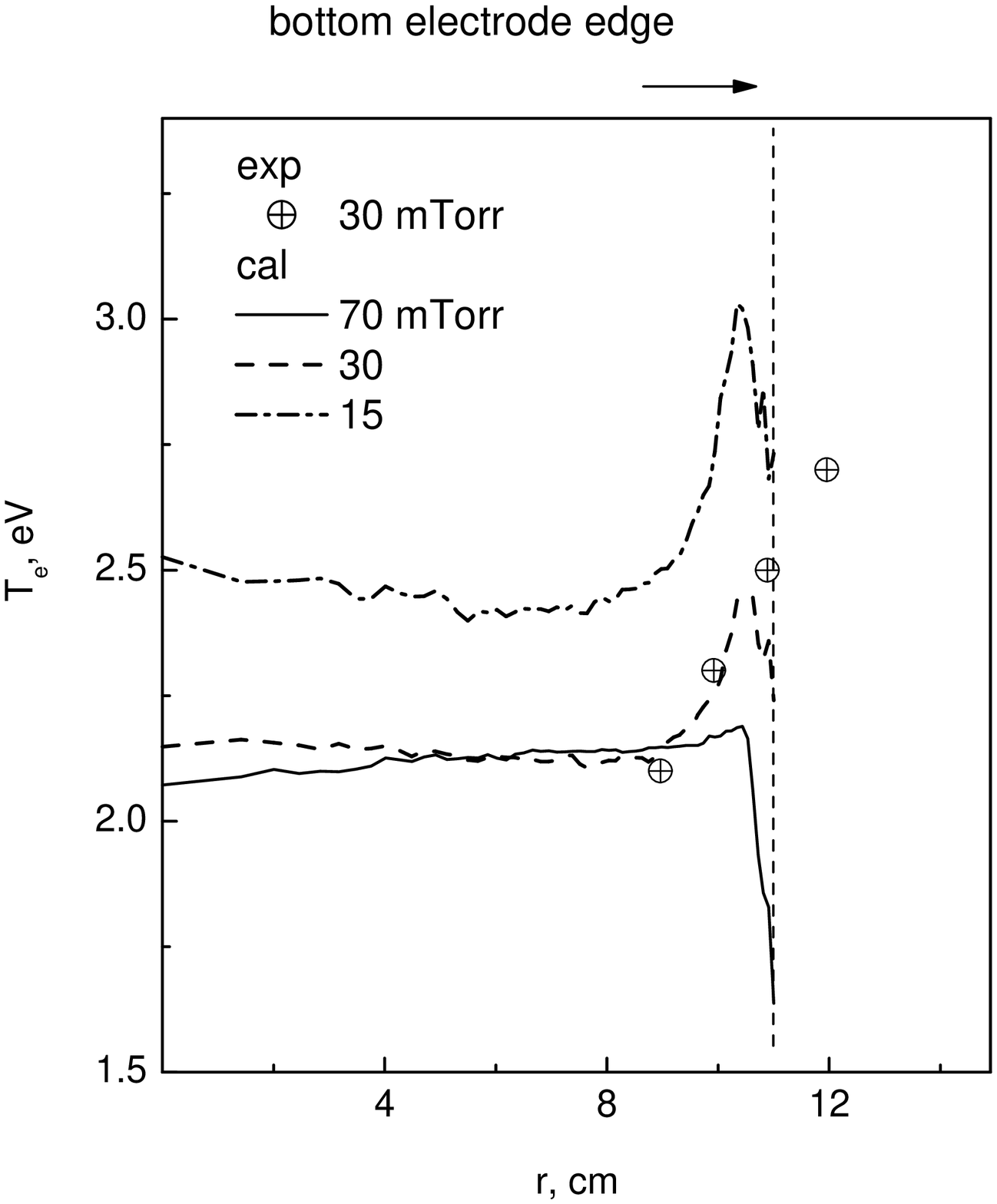}
\caption{Radial distributions of the electron temperature in chamber A
for P=30 mTorr (symbols, experiment) and in chamber B for P=15, 30, and
70 mTorr (curves, calculation).}
\label{ene_all}
\end{figure}

Figure \ref{ene_all} shows the radial profiles of the electron temperature
$T_e$=2/3$\epsilon_e$ at the center of the discharge gap for different
gas pressures. 
The mean electron temperature at $P=30$ mTorr, which was
measured near the electrode edge, agrees well with the numerical data.

The numerical and experimental data in Figs. \ref {den_all} and 
\ref {ene_all} are in good agreement and show that the discharge mode and its
parameters remain almost the same in both chambers. Because of the discharge
asymmetry, however, the potential drop in chamber B is substantially greater
near the grounded electrode, which makes it possible to form a high-energy
ion flux to the electrode and to control the distribution function of the ions.
  
\section{Ion flux to the electrode }
\label{ionflux}
\begin{figure}[h]
\includegraphics[width=0.8\linewidth]{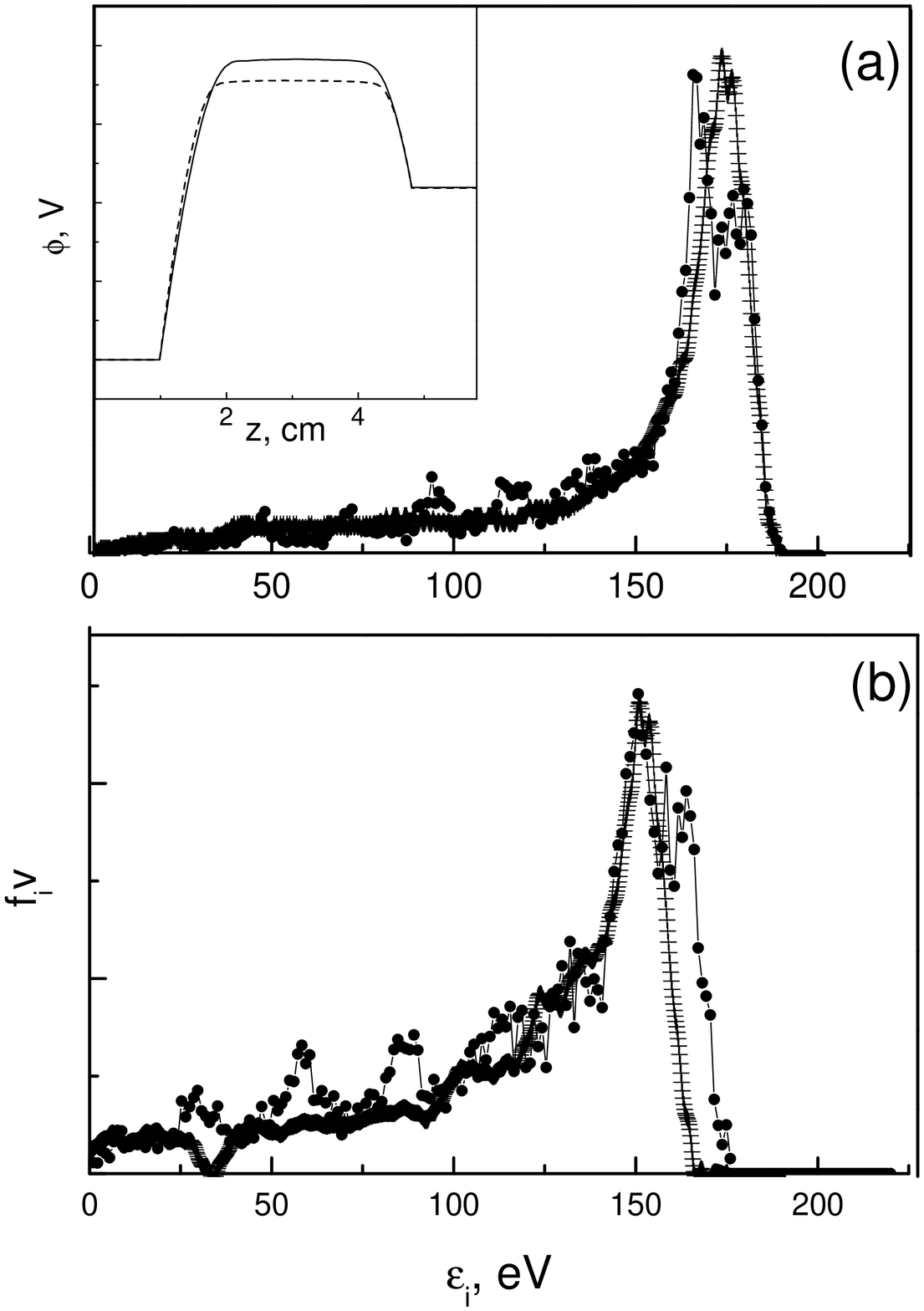}
\caption{
Ion flux distribution functions for 15 mTorr, $U_0$=286 V (à)
and 30 mTorr, $U_0$=260 V (b) in chamber B (the experimental and numerical
data are shown by crosses and circles, respectively). The inset shows the
potential distribution at r=0 for P=15 mTorr (solid curve) and P=30 mTorr 
(dotted curve).}
\label{idf_exp1}
\end{figure}

The experiment in chamber B was aimed at measuring the maximum energy of
ions on the electrode $E_{max}$ and the energy of ions $ E_p$ corresponding
to the maximum value of the ion distribution function (IDF) in terms of energy. 
As the value of the applied voltage cannot be accurately determined in the 
experiment, we varied the applied voltage in our calculations to obtain an IDF
close to the experimental curve. Figure \ref{idf_exp1} shows the measured
and calculated dimensionless IDFs at the center of the bottom electrode for
$P=15$ and 30 mTorr in chamber B. 
The measured and calculated plasma parameters
for the variants shown in Fig. \ref{idf_exp1} are summarized in Table \ref{table}.
The input power in the experiment was 14 W for $P=15$ mTorr
and 19.2 W for $P=30$ mTorr.

At gas pressures of 15 and 30 mTorr, the ions experience several collisions
as they cross the sheath; therefore, the ion energy corresponding to 
the IDF peak is $E_{p} < E_{max}$. The maximum energy of the ions obtained in
the experiment is $E_{max}$=189 eV at $P=15$ mTorr and $E_{max}$=191 eV
at $P=30$ mTorr. It is seen from Fig. \ref{idf_exp1} that the experimental
and calculated IDFs are in good agreement at the voltage $U_0$=285 V for 15 mTorr
and at $U_0$=260 V for 30 mTorr. The inset in Fig. \ref{idf_exp1} shows the
calculated distribution of the potential  $\phi $ at $r=0$. The self-bias
voltage is $U_{bias}$=108 V for both values of the gas pressure. The maximum
value of the plasma potential is $\phi_p$=191 V for $P=15$ mTorr and  170 V 
for $P=30$ mTorr. The maximum ion energy $E_{max}$ is approximately equal
to the averaged over 10 rf cycles plasma potential with respect to the
grounded electrode, because the characteristic time of the ion crossing
 the sheath is much greater than the discharge period.

\begin{figure}[h]
\includegraphics[width=0.85\linewidth]{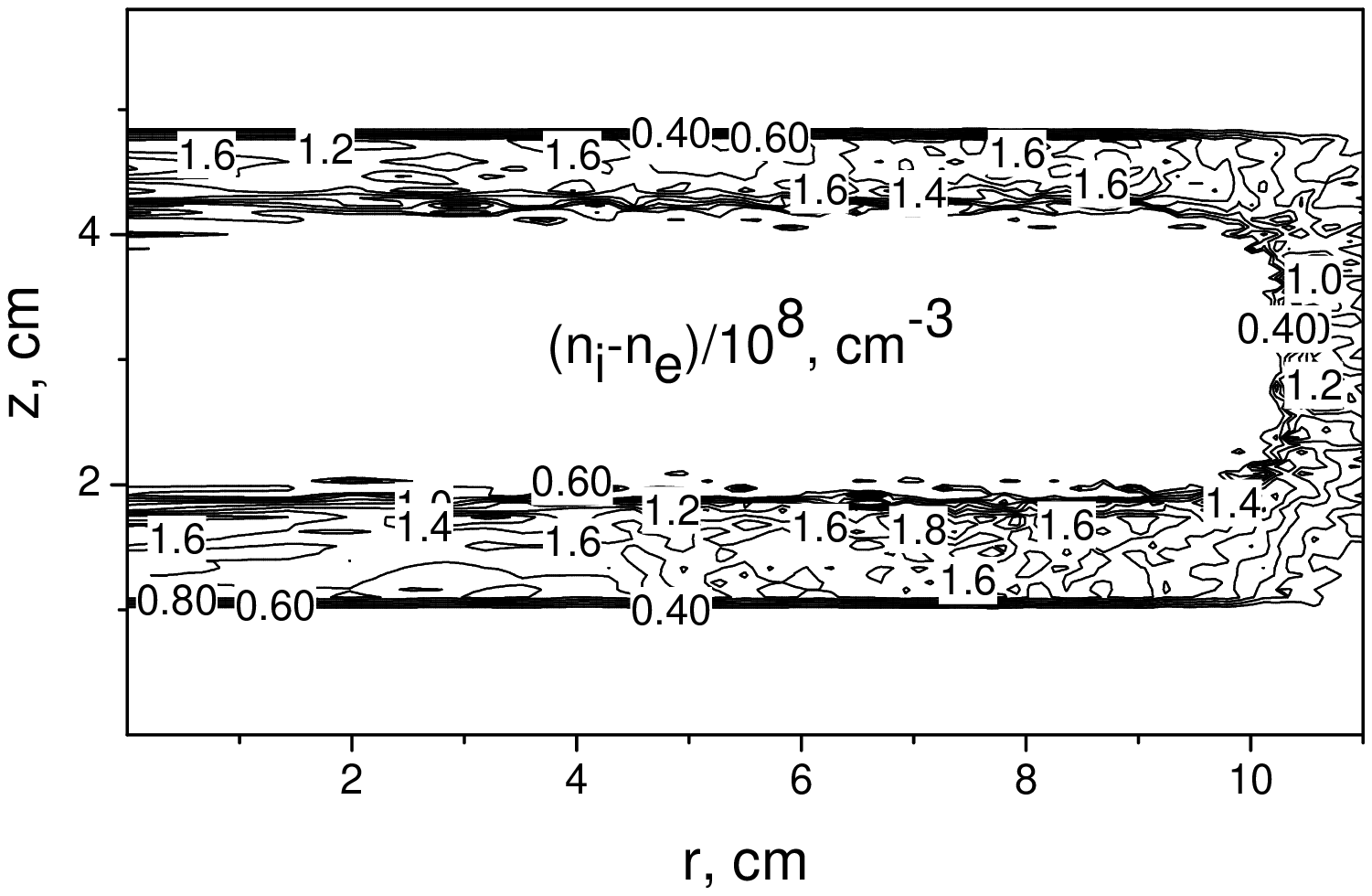}
\caption{Distribution of the positive charge for P=30 mTorr and $U_0$=260 V.} 
\label{char2D}
\end{figure}

The distribution of the charge characterizing the sheath width 
near the electrodes and chamber walls is shown in Fig. \ref {char2D}.
In the case considered, the sheath width is 1 cm for the grounded
electrode and approximately 0.5 cm for the surface with applied voltage.
At the pressure of 30 mTorr, the mean free path of ions with respect to
the charge exchange collisions with neutral atoms is approximately 2 mm; 
therefore, the ions
participate in several collisions when crossing the electrode sheath.

\begin{figure}[h]
\includegraphics[width=0.6\linewidth]{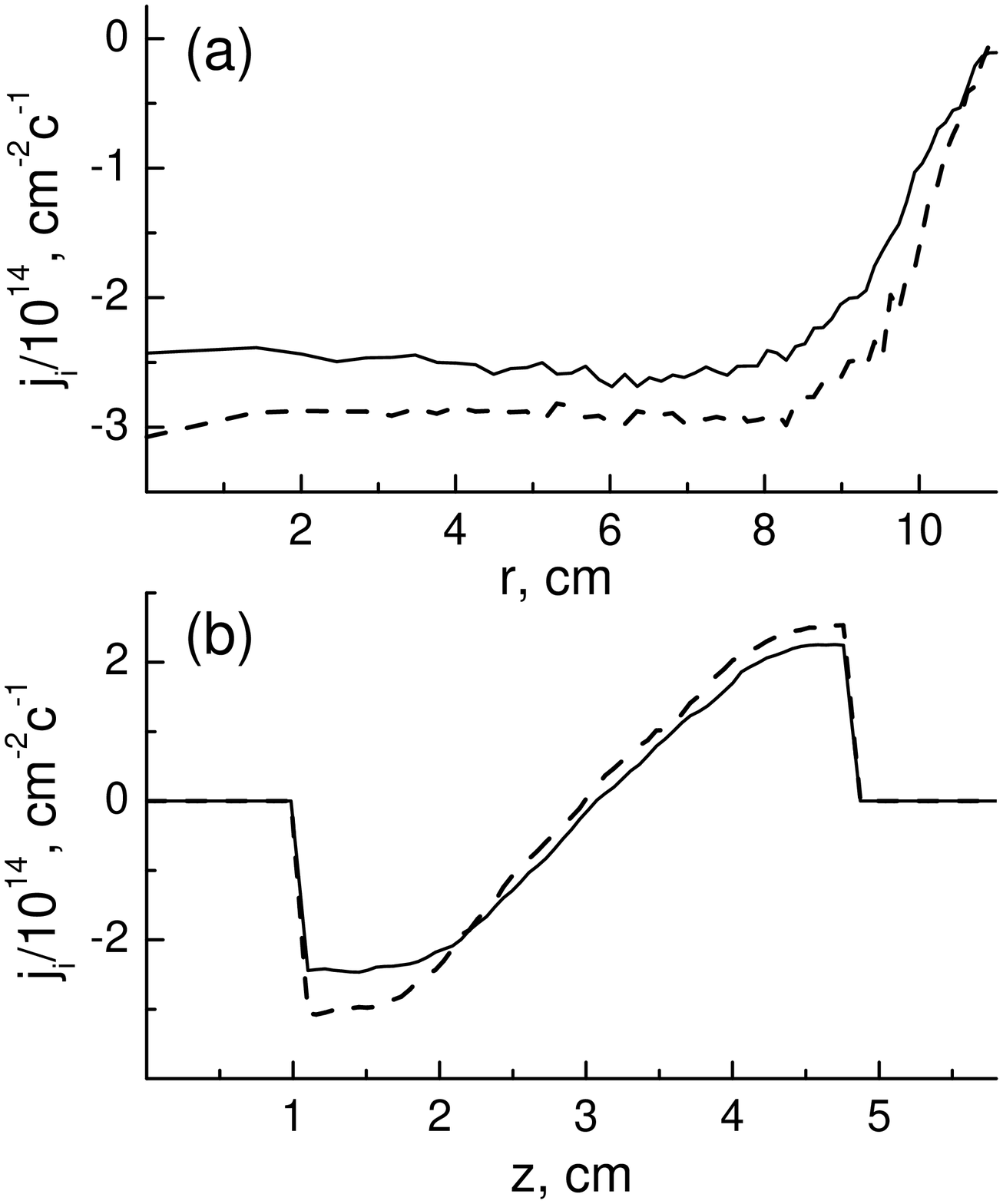}
\caption{
 Distributions of the ion flux along the bottom electrode (à)
and between the electrodes at r=0 (b) for 15 mTorr, $U_0$=286 V (solid curve)
and 30 mTorr, $U_0$=260 V (dotted curve).
}
\label{flux}
\end{figure}

Let us consider the ion flux distribution along the electrode surface.  
Figure \ref{flux} shows the distribution of the ion flux along the
surface of the bottom electrode and between the electrodes at $r=0$. 

The ion flux along the bottom electrode remains almost unchanged at 
$0<r<8$ cm and then decreases as the electrode edge is approached
(see Fig. \ref{flux}(a)). It is of interest to note that the ion flux
onto the top electrode is little different from the ion flux onto
the bottom electrode, though the ion energy on the top electrode is
considerably lower.
  
\section{Self-bias voltage for various discharge geometries} 
\label{poten}
\begin{figure}[h]
\includegraphics[width=0.6\linewidth]{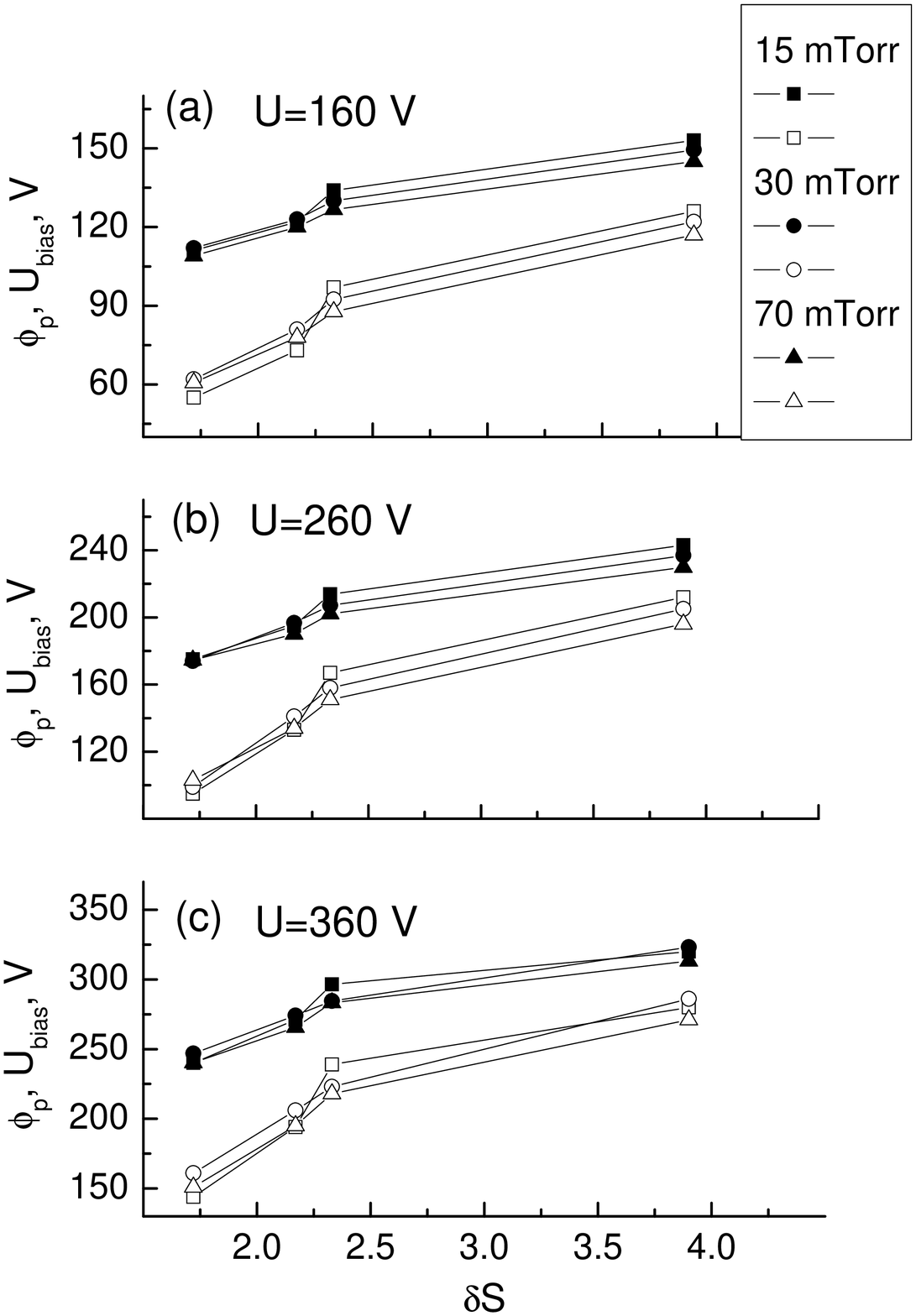}
\caption{
 Plasma potential relative to the grounded electrode (filled symbols)
and self-bias voltage (open symbols) for the voltage amplitude $U_0=160$ V (a), 
$U_0=260$ V (b), and $U_0=360$ V (c) for P=15, 30, and 70 mTorr.}
\label{geom}
\end{figure}

Let us consider the self-bias voltage in chambers of different geometries.
Figure \ref{geom} shows the calculated plasma potential relative to the 
grounded electrode $\phi_p$ and the self-bias voltage $U_{bias}$ 
as a function of the ratio of the areas of the driven and grounded
electrodes $\delta S$. 
The HFC discharge was calculated for chamber B
with different parameters: 
 1) R=11 cm, d=3.8 cm, and $\delta S$=1.72;
 2) R=7 cm, d=3.8 cm, and $\delta S$=2.17;
 3) R=11 cm, d=7 cm, and $\delta S$=2.33;
 4) R=6 cm, d=8 cm, and $\delta S$=3.9.
The bottom electrode and the chamber wall are separated by a 0.5-cm gap.
It is seen in Fig. \ref{geom} that the plasma potential increases with
increasing relative area of the driven electrode. In turn, 
$U_{bias}$ increases faster with $\delta S$ and approaches the plasma
potential value at $\delta S>$4.6. A decrease in the gas pressure leads to
an insignificant increase in the plasma potential relative to the
grounded electrode and the self-bias voltage. 

\begin{figure}[h]
\includegraphics[width=0.65\linewidth]{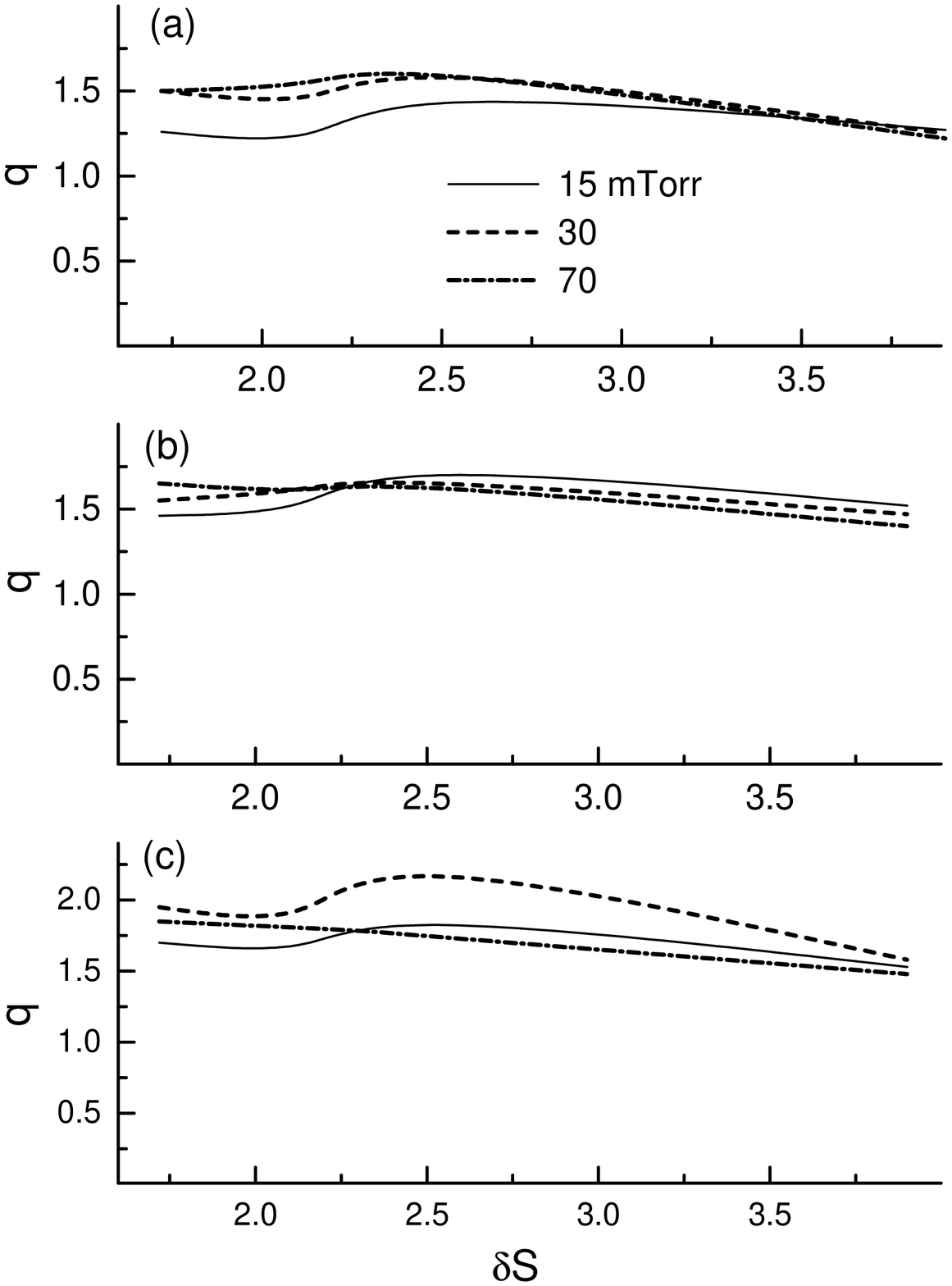}
\caption{Scaling exponent for the ratio 
$\phi_p/(\phi_p-U_{bias})=\delta S^{q}$ 
versus $\delta S$ for $U_0$=160 V (a), $U_0$=260 V (b), and
 $U_0$=360 V (c).}
\label{factor_S}
\end{figure}

In our calculations, we studied the dependence of the voltage drop in
the electrode sheaths on the electrode area ratio. For a symmetric discharge,
the ratio of the areas of the driven and grounded electrodes is
$\delta S$=1, and the ratio $\phi_p$/($\phi_p-U_{bias}$) is also equal to
unity. Using the results in Fig. \ref {geom}, we obtain a scaling exponent $q$ 
for the expression relating the voltage drop on the electrode sheaths to
the area of the electrodes in an asymmetric HFC discharge
$$\phi_p/(\phi_p-U_{bias})=\delta S^{q}.$$
Figure \ref {factor_S} shows the behavior of the exponent $q$ for
different pressures and voltages. 
Note that the curve for $q$  obtained by
two-dimensional kinetic calculations is a nonmonotonic function, which has
a maximum at $\delta S$=2-3: $q_{max}$=1.7 for P=15 mTorr, $q_{max}$=1.7 for
P=30 mTorr, and $q_{max}$ varies from 1.8 to 2.1 for P=70 mTorr. The exponent
calculated previously \citep{Lieberman89} with the use of a spherical model
was $q$=2.21. The value of $q$ varying from 1.6 to 2.1 was also obtained in
\citep {Schweigert2010} by means of kinetic calculations of an asymmetric 
2-MHz capacitive discharge.

\begin{figure}[h]
\includegraphics[width=0.65\linewidth]{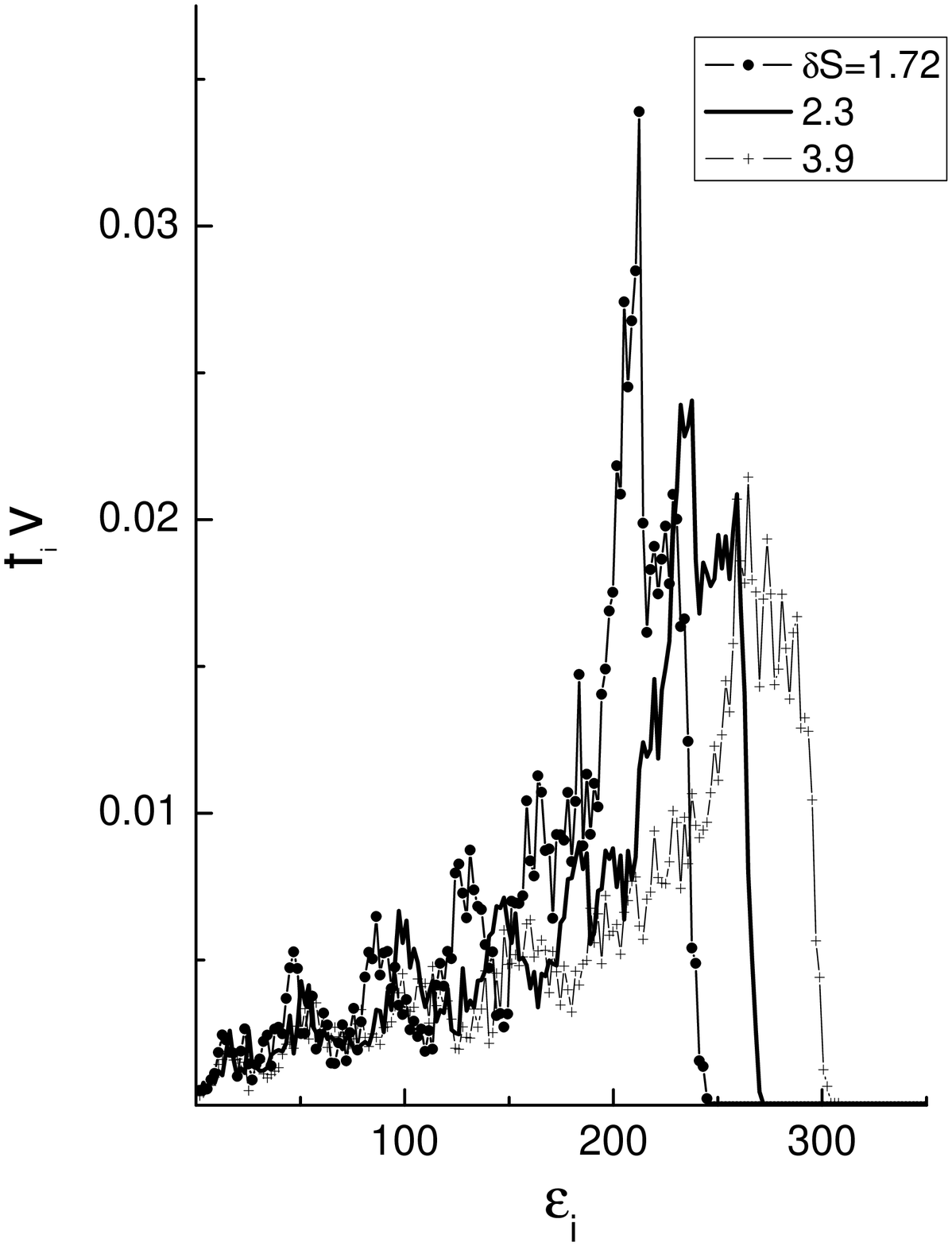}
\caption{Energy distribution function in the ion flux on the bottom electrode
in chamber B for different values of $\delta S$, Ð=30 mTorr, and $U_0$=360 V. }
\label{idf_360}
\end{figure}

Figure \ref{idf_360} shows the energy distribution functions in an ion flux
in chambers with different ratios of the areas of the driven and grounded 
electrodes $\delta S$. 
With increasing discharge asymmetry, the plasma
potential relative to the smaller electrode increases, which results in a greater
maximum energy of the ions.

\section{Conclusions}
\label{conclusion}
Formation of an ion flux in a low-pressure asymmetric HFC discharge was
studied in experiments and by means of kinetic numerical simulations.
Two-dimensional simulations were performed by the Particle-in-Cell method 
with collision sampling by the Monte Carlo method. In the experiment,
the ion flux was studied by an energy analyzer placed behind the grounded
electrode with an orifice in the middle. Several reactors with different
ratios of the areas of the driven and grounded electrodes were considered
to study the effect of the chamber geometry on the ion flux. The plasma potential
relative to the grounded electrode and, therefore, the maximum energy of 
the ions are demonstrated  to increase with increasing area ratio. The measured 
and calculated parameters of the plasma, such as the electron concentration 
and temperature, and also the ion energy distribution functions are in good
agreement.

\begin{acknowledgments}
The authors gratefully acknowledge the support of this work by the Russian
Foundation for Basic Research (grant No. $08-02-00833a$).
\end{acknowledgments}

\begin{center}
\begin{table}
\caption{Self-bias voltage $U_{bias}$, plasma potential $\phi_{p}$,
amplitude of applied voltage  $U_0$, concentration of electrons at the
center of the discharge gap $n_e$, and ion flux onto the bottom electrode
$j_i$ for different gas pressures.}
\begin {tabular}{lccccr} \hline
\hline
P(mTorr)&$U_{bias}$ (V)&$\phi_{p}$ (V)&$U_0$ (V)&$n_e$(cm$^{-3}$)&$j_i$ (s$^{-1}$cm$^{-2}$)
\\ \hline
&exp cal&exp cal&cal&cal&cal 
\\ \hline
$15 $&109 108&189 194 & 285 & 2.4$\times 10^{9}$  & 2.3$\times 10^{14}$ 
\\ \hline
$30 $&108 108&191 170 & 260 & 3.4$\times 10^{9}$  & 3.0$\times 10^{14}$ 
\\ \hline
\hline
\end{tabular}
\label{table}
\end{table}
\end{center}

\end{document}